\documentclass[aps,prb,reprint,superscriptaddress]{revtex4-1}

\usepackage[utf8x]{inputenc}
\usepackage{amsmath}
\usepackage[dvips]{graphicx}
\usepackage{amssymb}
\usepackage{color}
\usepackage{upgreek}

\renewcommand{\d}{\mathrm{d}}

\newcommand{\e}{\mathrm{e}}
\renewcommand{\i}{\mathrm{i}}
\newcommand{\varDeltaL}{\varDelta_\mathrm{L}}
\newcommand{\wL}{\omega_\mathrm L}
\newcommand{\wC}{\omega_\mathrm C}
\newcommand{\EC}{E_\mathrm C}

\begin{document}


\title{All-optical tailoring of single-photon spectra in a quantum-dot microcavity system}

\author{D. Breddermann}
\affiliation{Department of Physics and Center for Optoelectronics and Photonics Paderborn (CeOPP), Paderborn University, Warburger Strasse 100, 33098 Paderborn, Germany}
\author{D. Heinze}
\affiliation{Department of Physics and Center for Optoelectronics and Photonics Paderborn (CeOPP), Paderborn University, Warburger Strasse 100, 33098 Paderborn, Germany}
\author{R. Binder}
\affiliation{College of Optical Sciences, University of Arizona, Tucson, Arizona 85721, USA}
\author{A. Zrenner}
\affiliation{Department of Physics and Center for Optoelectronics and Photonics Paderborn (CeOPP), Paderborn University, Warburger Strasse 100, 33098 Paderborn, Germany}
\author{S. Schumacher}
\email{stefan.schumacher@upb.de}
\affiliation{Department of Physics and Center for Optoelectronics and Photonics Paderborn (CeOPP), Paderborn University, Warburger Strasse 100, 33098 Paderborn, Germany}
\affiliation{College of Optical Sciences, University of Arizona, Tucson, Arizona 85721, USA}

\date{\today}

\begin{abstract}
Semiconductor quantum-dot cavity systems are promising sources for solid-state based on-demand generation of single photons for quantum communication. Commonly, the spectral characteristics of the emitted single photon are fixed by system properties such as electronic transition energies and spectral properties of the cavity. In the present work we study cavity-enhanced single-photon generation from the quantum-dot biexciton through a partly stimulated non-degenerate two-photon emission. We show  that frequency and linewidth of the single photon can be fully controlled by the stimulating laser pulse, ultimately allowing for efficient all-optical spectral shaping of the single photon. 
\end{abstract}


\maketitle

\section{Introduction}

For applications in quantum communication, single photons of well defined and controllable spectral properties are needed \cite{knill2001scheme,kimble2008quantum,pan2012multiphoton,buckley2012engineered,Sollner2015}. To address this challenge, various flexible approaches for single-photon generation were introduced over the past years. On the one hand, single photon sources were realized where selected spectral parameters can be controlled during the creation process. In an ion-trap cavity system, it was demonstrated that the temporal structure of the emission can be imprinted 
by a driving optical field \cite{keller2004continuous}. Similarly, single photons with subnatural linewidth inherited from the exciting laser were generated via quantum-dot (QD) exciton resonance fluorescence \cite{matthiesen2012subnatural}. Also bandwidth and wavelength tunability through a whispering gallery mode resonator \cite{fortsch2013versatile} or through an enhanced two-photon Raman transition \cite{he2013indistinguishable,sweeney2014cavity} were realized, and pure frequency control with a strain-tunable QD structure \cite{zhang2015high}. Also control of single-photon properties after photon creation was realized, e.g., electro-optic  modulation was demonstrated for single photons emitted from atoms \cite{kolchin2008electro,specht2009phase} and spectral compression of photons via optical sum-frequency generation was achieved \cite{lavoie2013spectral}. 

\begin{figure}
   \centering
   \includegraphics[width=0.45\textwidth]{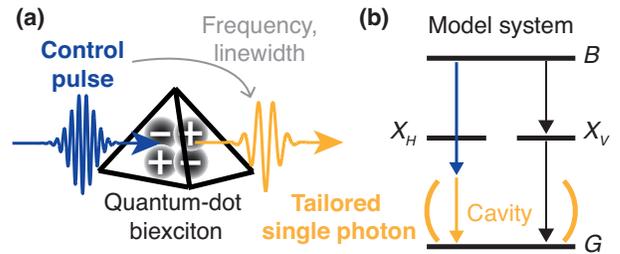}
   \caption{Sketch of emission scheme and spectral control. The external control pulse (blue arrow) triggers the first photon, the second one (yellow arrow) is spontaneously emitted. Panel (a) schematically illustrates the scheme.  Panel (b) shows the partly stimulated two photon transition between biexciton $B$ and ground state $G$ in the energy level diagram of the quantum-dot cavity system for the horizontal mode. The cavity (yellow vertical braces) enhances the second photon. It is off-resonant to the transitions of the biexciton-exciton cascade, indicated for the vertical mode starting from the biexciton (black arrows).}
    \label{fig:skizze}
  \end{figure}

Previous approaches have in common that the control only addresses certain properties of the single photon  and/or require additional non-optical control elements such as magnetic or strain fields. In this Article we analyze a relatively simple approach to gain access to all spectral properties of a single photon through the adjustment of a control-laser pulse.  Our general emission scheme has recently been introduced in the context of full and all-optical control on a single photon's polarization state \cite{heinze2015quantum}.  It is based on a non-degenerate enhanced two-photon emission \cite{yatsiv1968enhanced,braunlich1970detection} from the biexciton to the ground state in a commonly available solid-state system, a single semiconductor QD inside an optical microcavity \cite{michler2000quantum,strauf2007high,ota2011spontaneous,stock2013chip,jayakumar2013deterministic,hargart2016cavity,ardelt2016optical,ding2016demand,somaschi2016near}. As illustrated in Fig.~\ref{fig:skizze}, the first photon is triggered by a coherent control-laser field. The second photon is spontaneously emitted into a cavity mode when the QD relaxes to the ground state. In the present paper we demonstrate that, in addition to the polarization state, the external laser also controls the second photon's spectral shape, including emission frequency and linewidth (only limited by losses in the solid state system) because of two fundamental properties: (i) tunability of the frequency of the second photon by the control laser is a direct consequence of energy conservation in the two-photon process, (ii) the control-pulse induced photon emission is a purely stimulated and coherent process such that the emitted photon directly inherits the control pulse's linewidth. 
  
\section{Quantum-dot model and theory}

We model the QD as an effective four-level system in an electronic-configuration picture. To investigate the process of interest, we explicitly include the biexciton $B$, the ground state $G$, and the two linearly polarized excitons $X_{H(V)}$ interacting with horizontally (vertically) polarized light \cite{schumacher2012cavity}. Figure \ref{fig:skizze}(b) shows the electronic energy level scheme and the relevant optical transitions starting from the biexciton. The system Hamiltonian $H=H_0+H_\mathrm{I}$ contains the non-interacting Hamiltonian $H_0$ with the free electronic part and photons in two orthogonal cavity modes with polarizations $H$ and $V$, respectively. The light-matter interaction part of the Hamiltonian reads \cite{heinze2015quantum}
\begin{align}
 H_\mathrm{I}&=\sum_{i=H,V}\left[-g\left(P_{X_i,B}b_i^\dagger+P_{G,X_i}b_i^\dagger\right)+\mathrm{h.c.}\right]\notag\\
                      &\quad+\sum_{i=H,V}\left[\left(P_{X_i,B}\varOmega_i^\star(t)+P_{G,X_i}\varOmega_i^\star(t)\right)+\mathrm{h.c.}\right].
\end{align}
Here we define $P_{\alpha,\beta}\equiv|\alpha\rangle\langle\beta|$, which for $\alpha\neq\beta$ is the polarization operator between electronic states $\alpha$ and $\beta$, and the electronic occupation operator for $\alpha=\beta$. The photon operators $b_i^{(\dagger)}$ annihilate (create) a photon in the cavity mode with polarization $i$, with $i\in\{H,V\}$. The cavity-mode energy is $\hbar\omega_i$, its coupling strength to the QD excitations is given by $g$. The time dependent external classical laser field $\varOmega_i(t)$ couples to the QD transitions following the usual dipole selection rules for the photon-assisted transitions. We assume the exciton levels to be degenerate and choose the orthogonal cavity modes to have the same energy, $\hbar\omega_H=\hbar\omega_V\equiv\hbar\wC$. We note that even if degeneracy is lifted (in a range significantly smaller than the typical detunings and biexciton binding energy of several meV used below), the results are not significantly changed. The transition energies in our system are $E_{X_{H(V)}}-E_G=1.4$ eV, corresponding to 880 nm wavelength, and $E_B-E_{X_{H(V)}}=1.397$ eV, corresponding to a biexciton binding energy of 3 meV, typical of InGaAs-based QDs \cite{schumacher2012cavity}.
In the present work we study the emission for a QD system fully inverted to the biexciton state without any initial coherence or photons inside the cavity. Deterministic biexciton preparation can for example be achieved via two-photon Rabi-flopping \cite{heinze2015quantum,muller2014demand}. Optimum enhancement of the stimulated two-photon transition is achieved when control-pulse frequency $\wL$ (off-resonant to all single-photon transitions; here by few meV) and cavity frequency $\wC$ fulfill the energy-conservation condition $\hbar\wL+\hbar\wC\approx E_B-E_G$ (Fig 1(b)). In the following, for a fixed cavity frequency $\wC$, we tune the control frequency $\wL$ resulting in a detuning $\varDeltaL=\hbar\wC+\hbar\wL-(E_B-E_G)$ from the bare two-photon resonance condition. 

 \begin{figure}
   \centering
   \includegraphics[width=0.45\textwidth]{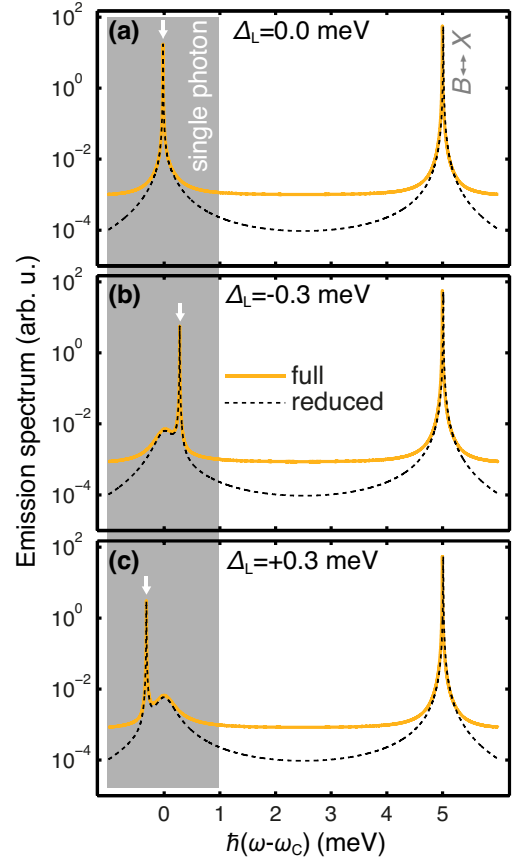}
   \caption{Controlling the single-photon emission frequency. Shown is the numerically computed CW emission spectrum (yellow solid line) for a control Rabi energy $\varOmega_0=0.25$ meV in comparison with the analytic result (black dashed line) for a detuning of (a) $\varDeltaL=0.0$ meV, (b) $\varDeltaL=-0.3$ meV and (c) $\varDeltaL=+0.3$ meV from the two-photon resonance condition (semi-logarithmically). The emission line of the single photon is marked by the white arrow and the biexciton-to-exciton emission by the label $B\leftrightarrow X$.}
   \label{fig:frequency}
  \end{figure}

To compute the temporal evolution of the system, we explicitly solve the equation of motion \cite{heinze2015quantum}, 
\begin{equation}
 \i\hbar\partial_t\rho=\left[H,\rho\right]+\left.\i\hbar\partial_t\rho\right|_\mathrm{cavity}+\left.\i\hbar\partial_t\rho\right|_\mathrm{pure}\,,
 \label{eq:EOM}
\end{equation}
for the system density operator $\rho$. We include a finite photon lifetime $\hbar/\kappa$ inside the cavity via the Lindblad form $\left.\i\hbar\partial_t\rho\right|_\mathrm{cavity}=\i\kappa/2\sum_{i=H,V}(2b_i\rho b_i^\dagger-b_i^\dagger b_i\rho-\rho b_i^\dagger b_i)$. Additionally, we introduce pure dephasing $\gamma$ of the electronic coherences by $\left.\i\hbar\partial_t\rho\right|_\mathrm{pure}=-\i\gamma/2\sum_{\alpha,\beta,\alpha\neq\beta}P_{\alpha,\alpha}\rho P_{\beta,\beta}$.  In a realistic parameter range, the influence of radiative losses on the single-photon emission spectrum is expected to be weak (see Appendix \ref{app:radloss} for a detailed discussion). The expectation value of any observable $O$ at time $t$ is then obtained by $\langle O\rangle(t)=\mathrm{tr}\{\rho(t)O\}$.  For all calculations presented here, we use $\hbar\wC=1.392$ eV ($5\,\mathrm{meV}$ below the resonance of the lowest energy single photon transition), $\gamma=\hbar/200\mbox{ ps}^{-1}$, $g=\hbar/50\mbox{ ps}^{-1}$ and $g/\kappa=0.04$. The cavity parameters correspond to a relatively low cavity qualtity of $Q\approx4,200$. These are typical and realistic parameters for our model system \cite{sweeney2014cavity,heinze2015quantum}. Numerically, Eq.~(\ref{eq:EOM}) is conveniently solved in the interaction picture after applying the unitary transformation $U(t)=\e^{\frac{\i}{\hbar} H_0t}$. Taking into account energy conservation  and the off-resonant character of cavity mode and control pulse, all operators are represented in a finite dimensional Fock space with a maximum photon number of two per polarization state. Here we apply the control pulse only in the horizontal optical mode such that the single photon of interest is also emitted into the $H$-mode of the cavity. To keep the notation as simple as possible, in the discussion below we drop the polarization-mode index of the exciton states, of the photon operators, and of the driving field.

To investigate the spectral properties of the cavity-enhanced QD emission, we compute the autocorrelation function $G(t,\tau)=\langle b^\dagger(t+\tau)b(t)\rangle$ at times $t$ and $\tau$ using the quantum regression theorem \cite{jahnke2012quantum}. For arbitrary excitations, e.g. pulsed scenarios, the time-dependent spectrum at frequency $\omega$ is given by \cite{eberly1977time,del2011generation,hargart2016cavity}
\begin{equation}
 S(t,\omega)=\Re\int_0^t\d t^\prime\int_0^{t-t^\prime}\d\tau\e^{-\i\omega\tau}G(t^\prime,\tau).
 \label{eq:EWS}
\end{equation}
In the stationary limit, e.g. continuous-wave (CW) excitations, and for long detection times $t$ the physical spectrum becomes proportional \cite{kira1999quantum,kabuss2010theory,hargart2016cavity} to the well-known Wiener-Khintchine sprectrum \cite{jahnke2012quantum}
\begin{equation}
 S_\mathrm{CW}(\omega)=\Re\int_0^\infty\d\tau\e^{-\i\omega\tau}G(\tau)
 \label{eq:WKS}
\end{equation}
where the autocorrelation function is stationary, $G(t,\tau)=G(\tau)$. 
Note that, in general, the linewidths and shapes produced by Eqs. \eqref{eq:EWS} and \eqref{eq:WKS}   are not directly comparable. However, we evaluate the spectrum Eq. \eqref{eq:EWS} only at times $t$ when the emission line for the single photon of interest is already fully established.

\section{Results}
\subsection{Frequency control}

First we show that the single-photon emission frequency can be tuned by varying the frequency of the control pulse/beam. For this purpose, we study the system response for varying control detuning $\varDeltaL$ in the CW limit, $\varOmega(t)=\varOmega_0\e^{-\i\wL t}$, and in the limit of weak excitations. This results in a quasi-stationary behavior of the biexciton population which is virtually not changing on the relevant time scale needed to fully establish the emission spectrum of the inverted system. Further below we show that in this limit additional and valuable analytical insight can be readily obtained into the spectral characteristics of the emission. The fundamental spectral features found in the CW limit are also recovered in the limit of control pulses of finite length. This is analytically discussed in more detail in Appendix \ref{app:CID} including radiative losses. Based on the full system dynamics described by Eq. \eqref{eq:EOM}, Fig. \ref{fig:frequency} shows the numerically computed CW emission spectra (yellow line) for three different control detunings $\varDeltaL$ in a semilogarithmic plot. Here, we focus on the spectral range including the region around the cavity energy (highlighted by the shaded area), where the single-photon emission takes place (marked by the small white arrow),  and the spectral range covering the spontaneous emission from the biexciton-to-exciton transition, off-resonant to the cavity mode (marked as $B\leftrightarrow X$). The emission from the exciton to ground state at $\hbar(\omega-\wC)=8$ meV is not included in the spectral range shown as it nearly vanishes in the quasi-stationary limit for a system initially inverted to the biexciton. For $\varDeltaL=0$ meV (panel (a)), we find that the single-photon emission occurs at the center of the cavity resonance. In panels (b) and (c), we observe that the line is blue shifted for negative and red shifted for positive detuning $\varDeltaL$ of the control beam following the energy conservation of the two-photon transition. For all three situations the biexciton-exciton line at $\hbar(\omega-\wC)=5$ meV remains uninfluenced by the control beam that is off-resonant to this transition.  Furthermore, the background emission at the cavity resonance is orders of magnitude weaker than the desired emission of the single photon. These results clearly demonstrate that, even in the weak-excitation regime, the control laser selectively drives the transition of interest and allows for all-optical control of the  frequency of the single photon emitted from the cavity. 

\begin{figure} 
   \centering
   \includegraphics[width=0.45\textwidth]{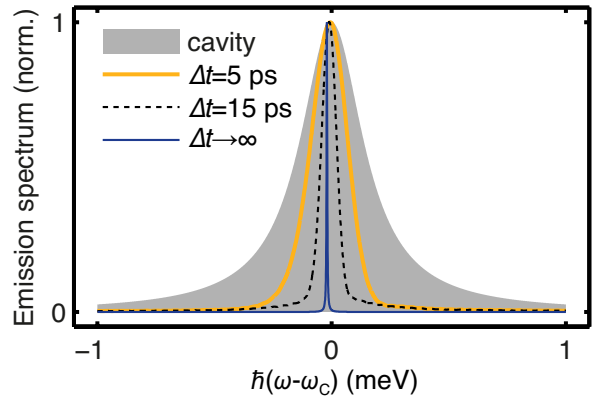}
   \caption{Controlling the single-photon emission linewidth. The computed single photon spectrum is shown for pulsed scenarios with parameters $\varOmega_0=0.75$ meV, $\varDelta t=5$ ps, $\varDeltaL=-0.09$ meV (yellow solid line) and $\varOmega_0=0.25$ meV, $\varDelta t=15$ ps, $\varDeltaL=0$ (black dashed line) together with the CW scenario from Fig. \ref{fig:frequency}(a) (blue solid line). Note, that the pulses have equal pulse areas. The cavity Lorentzian (shaded area) is shown for comparison. The spectra are each normalized to their respective peak maximum.}
   \label{fig:linewidth}
  \end{figure}
  
In the following we show that in the CW limit discussed above valuable additional insight can also be obtained analytically, helping to understand our observations in the numerical data of Fig.~2. To this end, we transform Eq.~(\ref{eq:EOM}) to the Heisenberg picture  and solve the equation of motion,
\begin{equation}
\i\hbar\partial_\tau G=-\left(\hbar\wC+\i\frac{\kappa}{2}\right)G+g\left[\varPi_{X,G}+\varPi_{B,X}\right]\,,
\end{equation}
for the stationary autocorrelation function $G(\tau)=\langle b^\dagger(\tau)b(0)\rangle$ analytically. Here we have defined $\varPi_{\alpha,\beta}(\tau)=\langle P_{\alpha,\beta}(\tau)b(0)\rangle$. As we are only interested in the single photon spectrum, we restrict the infinite set of equations of motion resulting from Eq.~(5), to the three central correlators $\varPi_{X,G}$, $\varPi_{B,X}$, and $\varPi_{B,G}$. For $\tau=0$, they are directly connected to the photon-assisted polarizations of the biexciton-exciton cascade, $\langle P_{X,B}b^\dagger\rangle$ and $\langle P_{G,X}b^\dagger\rangle$, and of the two-photon transition, $\langle P_{G,B}b^\dagger\rangle$, which is driven by the external control beam. The resulting {reduced model} obeys the equation of motion
  \begin{equation}
\i\hbar\partial_\tau \begin{pmatrix} \varPi_{X,G}\\ \varPi_{B,X} \\ \varPi_{B,G} \end{pmatrix}=-\begin{pmatrix} E_{X,G} & 0 & \varOmega(\tau) \\
    0 & E_{B,X} & -\varOmega(\tau)\\ \varOmega^\star(\tau) & -\varOmega^\star(\tau) &E_{B,G}\end{pmatrix}\begin{pmatrix} \varPi_{X,G} \\ \varPi_{B,X}\\ \varPi_{B,G} \end{pmatrix} 
\label{eq:EOMCorr}
\end{equation}
with $E_{\alpha,\beta}=E_\alpha-E_\beta+\i\gamma/2$.
  Using Eq. \eqref{eq:WKS} and defining $S_{\alpha,\beta}(\omega)=\int_0^\infty\d\tau\e^{-\i\omega\tau}\varPi_{\alpha,\beta}(\tau)$, the analytic CW emission spectrum then reads
\begin{equation}
 S_\mathrm{CW}(\omega)=\Re\left\{ \frac{\i\hbar G(0)+g\left[S_{X,G}(\omega)+S_{B,X}(\omega)\right]}{\hbar(\wC-\omega)+\i\kappa/2} \right\}\,.
 \label{eq:CWA}
\end{equation}
This analytic solution of our reduced model, is included in Fig. \ref{fig:frequency} (black dashed line) and shows excellent agreement with the full numerical result. It correctly reproduces all resonance features including peak heights, widths, and spectral positions of the single-photon line and of the biexciton-to-exciton emission. In the following we discuss the additional insight that can be obtained from Eq.~(7) into the details of the spectral features. The two fundamental components in Eq.~(7) are
\begin{align}
 S_{X,G}(\omega)&=\frac{\i\hbar\varPi_{X,G}(0)-\varOmega_0S_{B,G}(\omega+\wL)}{E_{X,G}-\hbar\omega}\,,\notag\\
 S_{B,X}(\omega)&=\frac{\i\hbar\varPi_{B,X}(0)+\varOmega_0S_{B,G}(\omega+\wL)}{E_{B,X}-\hbar\omega}\,.
\end{align}
These contain the spontaneous decay through the biexciton-exciton cascade via $\varPi_{X,G}(0)$ and $\varPi_{B,X}(0)$ at the respective emission frequencies. More importantly though for the present work, both also contain the control beam-induced source term responsible for the single-photon emission,
\begin{equation}
 S_{B,G}(\omega+\wL)=\frac{\i\hbar\varPi_{B,G}(0)+\varOmega_0\left[\frac{\i\hbar\varPi_{B,X}(0)}{E_{B,X}-\hbar\omega}-\frac{\i\hbar\varPi_{X,G}(0)}{E_{X,G}-\hbar\omega}\right]}{E_\mathrm{SP}(\omega)-\hbar\omega+\i\gamma/2}\,.
 \label{eq:SPL}
\end{equation}
The emission energy of the single photon can be identified as $E_\mathrm{SP}(\omega)=\hbar\wC-\varDeltaL -\varOmega_0^2\left[\frac{1}{E_{B,X}-\hbar\omega}+\frac{1}{E_{X,G}-\hbar\omega}\right]$. This energy linearly depends on the control beam detuning $\varDeltaL$ and thus can be conveniently tuned by changing the frequency of the control. We note that the strength of the single-photon contribution in Eq.~(8) linearly scales with the control beam amplitude. The additional control beam-induced energy shift will only be relevant for increasing control intensity as it scales with $\varOmega_0^2$. We note that for practical purposes this light-field induced shift can always be compensated by simply adjusting the control-beam frequency appropriately. In the limit of weak excitation, the emission energy is approximately given by $E_\mathrm{SP}-\hbar\wC\approx-\varDeltaL$ as expected from energy conservation. The resonance denominator of Eq.~\eqref{eq:SPL} shows that in the CW limit the linewidth is mainly given by the pure dephasing $\gamma$ (see also Appendix \ref{app:radloss}) which  is much narrower than the cavity's spectral width determined by the photon loss $\kappa$. The initial values, $G(0)=\langle b^\dagger b\rangle$ and $\varPi_{\alpha,\beta}(0)=\langle P_{\beta,\alpha}b^\dagger\rangle^\star$, are obtained analytically treating their equations of motion in the CW limit analogously to the equations of the correlators. Equations \eqref{eq:SPL}  and \eqref{eq:Gcav} can be used as a starting point for optimiziation of our single-photon source towards on-demand behaviour. A corresponding detailed numerical optimization study including further effects of quantitative importance (e.g., radiative losses and phonon-assisted processes) will be published elsewhere.

   \begin{figure}
   \centering
   \includegraphics[width=0.45\textwidth]{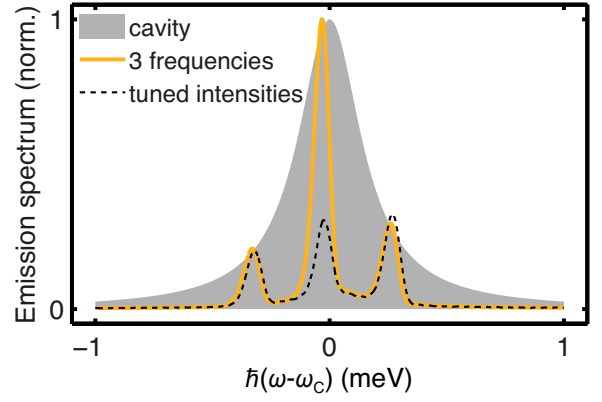}
   \caption{Spectral shaping of the single photon. Computed emission spectra for a Gaussian pulse  with $\varDelta t=15$ ps, containing three frequency components $\varDeltaL$ with equal intensities (yellow solid line) and tuned intensities (black dashed line). Details are given in the text. The spectra are normalized to the peak maximum of the yellow curve. The cavity line is shown for comparison (shaded area).}
   \label{fig:shaping}
  \end{figure}

\subsection{Spectral shape}

We now turn our attention back to scenarios with finite-length control pulses (taking into account the full system dynamics described by Eq.~(2)). As we show in Fig. \ref{fig:frequency}, the single-photon CW linewidth can be significantly narrower than the cavity line. To analyze the achievable linewidth more systematically, Fig. \ref{fig:linewidth} shows three cases with different control linewidth in comparison together with the cavity Lorentzian (shaded area). Starting from a Gaussian control pulse  $\varOmega(t)=\varOmega_0\e^{-\frac{(t-t_0)^2}{2\varDelta t^2}}\e^{-\i\wL t}$ (centered at time $t_0>0$) with $\varDelta t=5$ ps (yellow line), we increase the pulse length to $\varDelta t=15$ ps (black dashed line) up to the CW limit  $\varDelta t\rightarrow\infty$. We clearly observe that the spectral width of the single-photon line decreases with increasing length of the control. In all three examples, even for relatively short pulses, the single-photon emission line is narrower than the cavity line. In other words, for a cavity with relatively low quality the linewidth of the single-photon emission can be directly controlled with the duration of the control pulse. This is shown analytically for our reduced model in Appendix \ref{app:CID} also taking radiative losses into account.

Above we have shown that both frequency and linewidth of the single-photon emission can be tuned separately. Combining both mechanisms, also the overall spectral shape of the single photon can be designed using a simple pulse-shaping approach based on an appropriate superposition of control pulses with different detunings $\varDeltaL$, widths $\varDelta t$, and amplitudes $\varOmega_0$. As a first example, here we choose a superposition of three Gaussian control pulses with identical envelopes ($\varOmega_0=0.25$ meV, $\varDelta t=15$ ps) but different excitation frequencies corresponding to the detunings of $\varDeltaL=-0.3\,\mathrm{meV}, 0\,\mathrm{meV}, +0.3\,\mathrm{meV}$ of Fig. \ref{fig:frequency}. The resulting single-photon emission spectrum is depicted in Fig.~\ref{fig:shaping} (yellow line) together with the cavity Lorentzian (shaded area). We observe that the single-photon line is split into three separate frequencies. Their respective intensities follow the envelope of the cavity resonance. We repeat the calculation with the amplitude of the central control frequency at $\varDeltaL=0$ reduced to one half (black dashed line). Now the three spectral components exhibit similar shapes and intensities (black dashed line). The slight asymmetry of the spectral shape mainly stems from the light-induced energy shift in Eq.~\eqref{eq:SPL} and in its generalized form in Eq.~\eqref{eq:Gcav}. Our numerical and analytical results indicate that any spectral shape of the single photon that is compatible with Fourier transform and system parameters can be generated via simple pulse shaping of the control pulse, even in the presence of a cavity.

\section{Conclusions}

We have demonstrated a powerful approach to tailor the spectral properties of single photons emitted from commonly available semiconductor QD-cavity systems. Our approach offers all-optical and fully flexible control of the single-photon wavelength, linewidth and intensity at the moment of its creation. We also demonstrate single photon spectral shaping inside the cavity emission line. This work further paves the way towards a solid-state based on-demand source of tailored single photons.

\begin{acknowledgments}
 We gratefully acknowledge financial support from the DFG through the research centre TRR142 and doctoral training center GRK1464, from the BMBF through Q.com 16KIS0114, and a grant for computing time at $\mbox{PC}^2$ Paderborn Center for Parallel Computing. Stefan Schumacher further acknowledges support through the Heisenberg programme of the DFG.
 \end{acknowledgments}

\appendix
\section{Radiative decay}\label{app:radloss}
Besides photon losses $\kappa$ and pure dephasing $\gamma$ which are already included in the discussion in the main body of this article, also radiative losses $\delta$ from the excitons and the  biexciton can contribute to the system dynamics. Although this contribution is expected to be small (typically $\delta<1$ $\upmu$eV \cite{del2011generation,hargart2016cavity}) compared to the pure dephasing (typically several $\upmu$eV \cite{schumacher2012cavity,hargart2016cavity}), we discuss its  influence on the emission spectrum for completeness. Radiative decay can be modeled by introducing an additional Lindblad term \cite{heinze2015quantum} to the equation of motion Eq. \eqref{eq:EOM}, 
\begin{equation}
 \left.\i\hbar\partial_t\rho\right|_\mathrm{rad}=\frac{\i\delta}{2}\sum_{i}\sum_{\sigma_i}(2P_{\sigma_i}\rho P_{\sigma_i}^\dagger-P_{\sigma_i}^\dagger P_{\sigma_i}\rho-\rho P_{\sigma_i}^\dagger P_{\sigma_i}),
\end{equation}
for the transitions $P_{\sigma_i}$ with $\sigma_i=(G,X_i),(X_i,B)$ and $i=H,V$. This leads to modified loss constants $\varGamma_{\alpha,\beta}$ in the equations of motion  of the three fundamental correlators, $\varPi_{X_i,G}(t,\tau)=\langle P_{X_i,G}(t+\tau)b(t)\rangle$, $\varPi_{B,X_i}(t,\tau)=\langle P_{B,X_i}(t+\tau)b(t)\rangle$, describing the cascade, and $\varPi_{B,G}(t,\tau)=\langle P_{B,G}(t+\tau)b(t)\rangle$, describing the control-induced single-photon emission. These are defined as
\begin{equation}
 \i\hbar\partial_\tau\left.\varPi_{\alpha,\beta}(t,\tau)\right|_\mathrm{pure+rad}=-\i\varGamma_{\alpha,\beta}\varPi_{\alpha,\beta}(t,\tau)
\end{equation}
and are explicitly given as
\begin{equation}
  \varGamma_{X_i,G}=\frac{\gamma+\delta}{2},\quad \varGamma_{B,X_i}=\frac{\gamma+3\delta}{2},\quad \varGamma_{B,G}=\frac{\gamma+2\delta}{2}.
\end{equation}
In other words, $\varGamma_{\alpha,\beta}$ defines the natural emission linewidth of the corresponding transition. In particular, $\varGamma_{B,G}$ is the natural linewidth of the controlled single-photon emission which is observable in the emission spectrum in the CW limit and which in the case of small $\delta$ is mainly given by the pure dephasing $\gamma$. 

\section{Control-induced spectrum}\label{app:CID}

In this appendix we derive a generalization of Eq. \eqref{eq:SPL} for control pulses $\varOmega(t)=\varOmega_\mathrm{env}(t)\e^{-\i\wL t}$ with arbitrary envelope $\varOmega_\mathrm{env}$. This will lead to a deeper understanding of the influence of the QD-cavity design (including photon/radiative losses and pure dephasing) and the design of the control pulse on the fundamental properties of the single photon's spectrum. For this purpose, we rewrite Eq.~\eqref{eq:EOMCorr} for the correlators $\varPi_{\alpha,\beta}(t,\tau)=\langle P_{\alpha,\beta}(t+\tau)b(t)\rangle$. Going into the rotating frame, $\varPi_{\alpha,\beta}(t,\tau)=\e^{\i E_{\alpha,\beta}\tau/\hbar}\varPsi_{\alpha,\beta}(t,\tau)$ with $E_{\alpha,\beta}=E_\alpha-E_\beta+\i\varGamma_{\alpha,\beta}=\hbar\omega_{\alpha,\beta}$, the equation of motion reads
 \begin{equation}
\partial_\tau \begin{pmatrix} \varPsi_{X,G}\\ \varPsi_{B,X} \\ \varPsi_{B,G} \end{pmatrix}=\frac{\i}{\hbar}\begin{pmatrix} 0 & 0 & \varOmega_{1,3} \\
    0 & 0 &  \varOmega_{2,3}\\ \varOmega_{3,1}& \varOmega_{3,2}&0\end{pmatrix}\begin{pmatrix} \varPsi_{X,G} \\ \varPsi_{B,X}\\ \varPsi_{B,G} \end{pmatrix} \,.
\label{eq:EOMRWA}
\end{equation}
Here we have defined the control-pulse components
\begin{align}
 \varOmega_{3,1}(t,\tau) &= \varOmega^\star(t+\tau)\e^{-\i(\omega_{B,G}-\omega_{X,G})\tau},\notag\\
 \varOmega_{3,2}(t,\tau) &= -\varOmega^\star(t+\tau)\e^{-\i(\omega_{B,G}-\omega_{B,X})\tau},\notag\\
\varOmega_{1,3}(t,\tau) &= \varOmega(t+\tau)\e^{\i(\omega_{B,G}-\omega_{X,G})\tau},\notag\\
 \varOmega_{2,3}(t,\tau) &= -\varOmega(t+\tau)\e^{\i(\omega_{B,G}-\omega_{B,X})\tau}\,.
\end{align}
To solve the equation of interest, i.e., the third equation of Eq. \eqref{eq:EOMRWA}, we use the ansatz 
\begin{equation}
\varPsi(t,\tau)=\e^{\frac{\i}{\hbar}\int_0^\tau\d\tau' M(t,\tau')}\tilde\varPsi(t,\tau)
\end{equation} 
for the correlator vector $\varPsi=(\varPsi_{X,G},\varPsi_{B,X},\varPsi_{B,G})$ using the transformation matrix 
\begin{equation}
M(t,\tau)=\begin{pmatrix} 0 & 0 & 0 \\
    0 & 0 & 0\\ \varOmega_{3,1}(t,\tau) & \varOmega_{3,2}(t,\tau) &0\end{pmatrix}
\end{equation}
which fulfills $M(t,\tau)M(t,\tau')=0$ and trivially  $\left[M(t,\tau),M(t,\tau')\right]=0$. Considering the light field up to the second order, Eq.~\eqref{eq:EOMRWA} can be decoupled and solved analytically. We eventually obtain
\begin{equation}
 \varPsi_{B,G}(t,\tau)=\langle P_{B,G}b\rangle (t)\e^{-\frac{\i}{\hbar}\varSigma(t,\tau)\tau}
\label{eq:psibg}
\end{equation}
where we have identified the light-induced energy shift 
\begin{equation}
 \varSigma(t,\tau)=\frac{\i}{\hbar\tau}\int_0^{\tau}\d\tau'\int_0^{\tau'}\d\tau''\sum_{k=1}^2\varOmega_{3,k}(t,\tau'')\varOmega_{k,3}(t,\tau')
\end{equation}
which is second order in the field, as expected from Eq.~\eqref{eq:SPL}. With Eqs.~\eqref{eq:EOMRWA} and \eqref{eq:psibg}, the autocorrelation function is now fully determined via its equation of motion
\begin{equation}
 \partial_\tau G(t,\tau)=\frac{\i\EC}{\hbar} G(t,\tau)-\frac{\i g}{\hbar}\left(\varPi_{X,G}(t,\tau)+\varPi_{B,X}(t,\tau)\right)
\end{equation}
where we have introduced $\EC=\hbar\wC+\i\kappa/2$. Since we are only interested in the control-induced single-photon emission, we extract the contribution with frequency near the cavity frequency from the full result for $G(t,\tau)$ yielding
\begin{widetext}
\begin{align}
 \left .G(t,\tau)\right|_{\mathrm{Cavity}}&=\e^{\frac{\i}{\hbar}\EC\tau}\bigg(\langle b^\dagger b\rangle(t)+g\frac{\langle P_{X,G}b\rangle (t)}{E_{X,G}-\EC}+g\frac{\langle P_{B,X}b\rangle (t)}{E_{B,X}-\EC}\notag\\
          &\quad-\frac{\i g}{\hbar}\left[\frac{1}{E_{B,X}-\EC}-\frac{1}{E_{X,G}-\EC}\right]\langle P_{B,G}b\rangle (t)\e^{-\i\wL t}\int_0^\tau\d\tau' \e^{-\frac{\i}{\hbar}(\varDeltaL+\varSigma(t,\tau')+\i[\frac{\kappa}{2}-\varGamma_{B,G}])\tau'}\varOmega_\mathrm{env}(t+\tau')\bigg).
\label{eq:Gcav}
\end{align}
\end{widetext}
The corresponding emission spectrum can then be obtained using Eq. \eqref{eq:EWS}, which is more or less the Fourier transform with respect to the $\tau$-coordinate. For improved readability, we avoid to give its explicit form here and discuss its fundamental properties in the time domain represented by Eq. \eqref{eq:Gcav}.
The first line describes the background emission caused by the total photon emission $\langle b^\dagger b\rangle(t)$ and the contributions from the biexciton-exciton cascade $\langle P_{X,G}b\rangle (t)$, $\langle P_{B,X}b\rangle(t)$. The second line describes the control-induced single-photon emission: its strength is determined by the QD-photon coupling $g$, the difference of the reciprocals of the cavity-detuned transition energies -- which is proportional to the biexciton-binding energy --, and the pulse amplitude. Its width and shape are either defined by the width and shape of the pulse envelope $\varOmega_\mathrm{env}$ following Fourier-transform rules or, in the CW limit, by the QD linewidth $\varGamma_{B,G}$ consisting of the pure dephasing $\gamma$ and the radiative decay $\delta$. Furthermore, the spectral position within the cavity line is controllable via $\varDeltaL$. Note that the light-induced energy shift $\varSigma$ introduces an additional detuning for increasing pulse intensity which can be compensated by an appropriate choice of $\varDeltaL$.
\bibliography{Bibliography.bib}

\end{document}